\documentclass[journal,10pt]{IEEEtran}
\usepackage{blindtext, graphicx}
\usepackage{cite}
\usepackage{array}
\usepackage{multirow}
\usepackage{amssymb,mathtools,bm}
\usepackage{color}
\usepackage{algorithm}
\usepackage[algo2e]{algorithm2e}
\usepackage{subfigure}
\usepackage{amsmath}
\usepackage{lipsum, cuted}

\newcolumntype{P}[1]{>{\centering\arraybackslash}p{#1}}
\def\BibTeX{{\rm B\kern-.05em{\sc i\kern-.025em b}\kern-.08em
    T\kern-.1667em\lower.7ex\hbox{E}\kern-.125emX}}
\begin{document}

\title{Machine Learning Aided Holistic Handover Optimization for Emerging Networks}

\author{\IEEEauthorblockN{Muhammad Umar Bin Farooq\IEEEauthorrefmark{1},
Marvin Manalastas\IEEEauthorrefmark{1},
Syed Muhammad Asad Zaidi\IEEEauthorrefmark{1}, 
\\Adnan Abu-Dayya\IEEEauthorrefmark{2}, and
Ali Imran\IEEEauthorrefmark{1}}\\
\IEEEauthorblockA{\IEEEauthorrefmark{1}AI4Networks Research Center, School of Electrical \& Computer Engineering, University of Oklahoma, USA}
\IEEEauthorblockA{\IEEEauthorrefmark{2}Department of Electrical Engineering, Qatar University, Doha Qatar\\
Email: \{umar.farooq,marvin, asad, ali.imran\}@ou.edu, adnan@qu.edu.qa }}

\markboth{Accepted in IEEE International Conference on Communications (ICC) 2022}
{}
\maketitle

\begin{abstract}
In the wake of network densification and multi-band operation in emerging cellular networks, mobility and handover management is becoming a major bottleneck. The problem is further aggravated by the fact that holistic mobility management solutions for different types of handovers, namely inter-frequency and intra-frequency handovers, remain scarce. This paper presents a first mobility management solution that concurrently optimizes inter-frequency related A5 parameters and intra-frequency related A3 parameters. We analyze and optimize five parameters namely A5-time to trigger (TTT), A5-threshold1, A5-threshold2, A3-TTT, and A3-offset to jointly maximize three critical key performance indicators (KPIs): edge user reference signal received power (RSRP), handover success rate (HOSR) and load between frequency bands. In the absence of tractable analytical models due to system level complexity, we leverage machine learning to quantify the KPIs as a function of the mobility parameters. An XGBoost based model has the best performance for edge RSRP and HOSR while random forest outperforms others for load prediction. An analysis of the mobility parameters provides several insights: 1) there exists a strong coupling between A3 and A5 parameters; 2) an optimal set of parameters exists for each KPI; and 3) the optimal parameters vary for different KPIs. We also perform a SHAP based sensitivity to help resolve the parametric conflict between the KPIs. Finally, we formulate a maximization problem, show it is non-convex, and solve it utilizing simulated annealing (SA). Results indicate that ML-based SA-aided solution is more than 14x faster than the brute force approach with a slight loss in optimality.

\end{abstract}


\IEEEpeerreviewmaketitle

\section{Introduction}

The ever-increasing user demands for high data rates, large number and variety of connected devices, and the low latency-dependent use cases are some of the major drivers for the evolution of cellular networks. To meet these requirements, one of the most formidable approaches for the emerging networks is a shift towards denser and heterogeneous deployment containing different types of base stations (BS) operating at motley of frequency bands \cite{6963801}. For instance, 4G only exploits sub-6GHz frequency bands but the most recent 5G-New Radio (NR) also operates on mmWave spectrum alongside the sub-6GHz range. This range of frequency band is expected to further broaden in the upcoming 6G networks with the utilization of the THz band. However, the migration towards dense heterogeneous networks unravels unprecedented challenges specifically on energy efficiency \cite{farooq2018user} and mobility management. The mobility related challenges arise from the inevitable explosion in handover (HO) occurrences and the upsurge in signaling overhead.

Mobility management in cellular networks plays a pivotal role in ensuring an optimal quality of service for users. This is due to the direct impact of HO settings on KPIs such as HO success rate (HOSR), retainability, cell utilization, and throughput to name a few. This relationship between HO settings and KPIs are presented in a comprehensive survey on the mobility management in 5G and emerging networks \cite{zaidi2020mobility}. To optimize HO, the current industrial practice relies on vendor defined gold standards (GS) and on the human experience-based tuning of HO related configuration and optimization parameters (COPs). GS is based on a one-model-fits-all evaluation, which is not sufficient to cover the vast and distributed nature of the emerging networks. In addition, the human intervention in the manual tuning of COPs is not suitable for rapidly changing network conditions, not to mention its vulnerability to human errors. To reduce human dependency, self organizing networks (SON) induce some level of automation in the optimization process. One specific SON module known as mobility robustness optimization (MRO) optimizes a limited number of HO-related COPs based on past HO records. However, this approach relies on semi-automated hit-and-trial in most cases, which makes it reactive in nature. To meet the needs of 5G and future 6G networks, the pressing need to depart from a semi-automated approach towards a fully automated HO management has never been evident.   

A proactive and fully automated HO management starts with an efficient HO optimization framework designed to quantify the relationship between HO-related COPs and affected KPIs. However, despite the recent efforts to analytically model the COP-KPI relationship \cite{vasudeva2016analysis}, \cite{xu2017modeling}, tractable models for several COP-KPI associated with HO are not possible due to diverse user mobility and varying network dynamics. To overcome this challenge, data driven models are an effective alternative. These models are capable of orchestrating functions, which can map out KPI vis-a-vis HO-related COPs. However, these models are extremely sensitive to data sparsity and poor data representation. In cellular networks, gathering rich and representative data for training data driven models is considered as a bottleneck due to privacy concerns and possible network impairment while carrying out trials on a wide range of COP combinations in a live network. In this backdrop, utilizing other methods for data generation, such as simulators, can be beneficial.

\subsection{Related Work}
\label{sec:relatedWork}

The importance of mobility management is manifested by the unwinding interest of both the research community and industry players in optimizing HO. Recent work in mobility management involves HO-related parameter optimization \cite{castro2017optimization,silva2018adaptive,joseph2020big,nguyen2017mobility,9348101,umar2021data} and proposing new methods for performing HO \cite{qi2021social}.

Authors in \cite{castro2017optimization} categorized users in signal strength-based clusters and tuned time to trigger (TTT) and offset of event A3. The goal of the study was to improve spectral efficiency and data rates of edge users. On the other hand, a fuzzy logic-based scheme utilizing user velocity to adapt the HO margin (HOM) of event A3 was proposed in \cite{silva2018adaptive}. The proposed scheme aimed to reduce the HO failure ratio, the number of HO, and ping-pongs. Authors in \cite{joseph2020big} proposed a channel individual offset (CIO) tuning algorithm for event A3. The algorithm tuned CIO for each cell-pair and commercial LTE testing of algorithm reduced radio link failures (RLF). To reduce RLF, authors in \cite{nguyen2017mobility} used three COPs (TTT, offset of A3 and CIO) and developed a distributed algorithm. While \cite{castro2017optimization,silva2018adaptive,joseph2020big,nguyen2017mobility} delved in improving HO of cells operating on a single frequency band called intra-frequency HO, authors in \cite{9348101,umar2021data} attempted to optimize HO between cells with different frequency bands called inter-frequency HO. They used threshold1, threshold2, and TTT of event A5 to optimize reference signal received power (RSRP) and HOSR. Moreover, they exploited genetic algorithm for optimization and showed a faster convergence compared to the brute force method. Meanwhile, a new HO scheme utilizing the signal strength, available bandwidth, and the sojourn time was presented in \cite{qi2021social}. A social long short-term memory (LSTM) was used for the sojourn time prediction. The proposed HO scheme reduced the number of HO and ping-pongs.

The aforementioned works on HO management either optimize intra-frequency HO using event A3 parameters or inter-frequency HO using event A5 parameters. However, the current approach of partial optimization of intra and inter-frequency HO parameters is not optimal because there exists a strong interdependence between event A3 and event A5 parameters. Motivated by these shortcomings, we present a holistic mobility management framework, which simultaneously optimizes A3 and A5 parameters to improve multiple KPIs. On top of the HOSR, this multi-objective optimization framework jointly optimizes two other KPIs namely edge user RSRP and load distribution among different frequency bands.

\subsection{Contributions}
The main contributions of the work are listed below:
\begin{enumerate}
    \item This paper presents a holistic inter and intra-frequency HO parameter optimization framework utilizing five mobility COPs namely, A5 TTT, A5 threshold1, A5 threshold2, A3 TTT, and A3 offset. We show that optimizing inter-frequency and intra-frequency in silos is not optimal as there exists a strong coupling between them for several KPIs. To the best of the authors' knowledge, there does not exist any framework for simultaneous optimization of both inter and intra-frequency HO.
    \item We formulate and solve a multi-objective optimization problem to achieve the optimal values of the 5 COPs that jointly maximize edge user RSRP and HOSR while balancing the load distribution between different frequency bands. 
    In the absence of tractable analytical models due to system level complexity, we utilize data driven modeling to quantitatively mine the COP-KPI relationship. Evaluations of several state-of-the-art models reveal that XGBoost outperforms other methods in predicting edge user RSRP and HOSR while random forest has the best performance for load balancing.
    \item To resolve the objective conflict between different KPIs, we leverage shapley additive explanation (SHAP) analysis. The insights drawn from the SHAP analysis are valuable especially for network operators to understand and deal with the discord that usually arises with multi-objective KPI optimization.
    \item Finally, we establish the non-convexity of the optimization problem and leverage simulated annealing to solve the problem. Results show that the SA-aided data driven optimization converges around 14 times faster compared to the brute force approach. The faster convergence is particularly useful for rapidly changing network conditions and dynamics.
\end{enumerate}

The rest of the paper is organized as follows: we present a system model including event A3 and A5, problem formulation, and data generation in Section \ref{sec:systemModel}. Section \ref{sec:impact} presents the behavior of KPIs with varying COPs. Meanwhile, Section \ref{sec:ML} describes the performance of ML algorithms and SHAP sensitivity analysis. The objective function optimization is discussed in Section \ref{sec:optimization} and Section \ref{sec:conc} concludes the paper.

\section{System Model}
\label{sec:systemModel}
In this section, we describe 3GPP-based HO events A3 and A5 as well as the optimization KPIs such as edge users RSRP, HOSR, and load balancing. We then formulate the optimization problem followed by the data generation process.

\subsection{Event A3-based Intra-frequency Handover}
Event A3-based intra-frequency HO is triggered when the RSRP of a user $u$ from target gNB becomes higher than the RSRP of the user from the serving gNB by an offset. Note that here, both the source and target gNB operate using a similar frequency band. If the following condition is maintained for a certain time called time to trigger, $A3_{TTT}$, A3-based HO is triggered:
\begin{equation}
    M_u^t + O_{s,t} - A3_{hyst} > M_u^s + A3_{off}\\
\label{eq:A3_entering}
\end{equation}
where $M_u^t$ and $M_u^s$ is the RSRP from target $t$ and serving gNB $s$ to the user $u$, respectively, $O_{s,t}$ is cell specific offset from serving gNB to target gNB also known as CIO, while $A3_{hyst}$ and $A3_{off}$ are the hysteresis and offset of event A3, respectively. 

\subsection{Event A5-based Inter-frequency Handover}
Event A5-based inter-frequency HO is triggered when the RSRP of a user from serving gNB decreases below a threshold, i.e., threshold1, and the RSRP of the same user from a target gNB increases above another threshold, i.e., threshold2. Unlike event A3-based HO, the source and target gNB in this case operates on different frequency bands. HO using A5 is triggered when the following conditions remain satisfied for a $A5_{TTT}$.
\begin{equation}
\begin{gathered}
    M_u^s + A5_{hyst} < A5_{th1}\\
    M_u^t + O_{s,t} - A5_{hyst} > A5_{th2}
\end{gathered}
\label{eq:A5_entering}
\end{equation}
where $A5_{hyst}$, $A5_{th1}$ and $A5_{th2}$ are the hysteresis, threshold1 and threshold2 for event A5, respectively. 



\subsection{Problem Formulation}
We begin the problem formulation with the presentation of path loss model. In this paper, we model the path loss between the users and gNB as a close-in (CI) dual slope path loss model \cite{sun2015path}. The dual slope path loss equation for the CI model is expressed as follows:
\begin{equation}
\small
  PL_{Dual}^{CI}(d_u^s) =
    \begin{cases}
      -FSPL(1m) - 10n_1log_{10}(d_u^s) & \text{for } d_u^s \leq d_{th}\\
       -FSPL(1m) - 10n_1log_10(d_u^s)\\
       - 10n_2log_{10}(d_u^s/d_{th}) & \text{for } d_u^s > d_{th}
    \end{cases}       
    \label{eq:PL}
\end{equation}
where PL is the path loss in dB, $d_u^s$ is the 3D distance between the serving gNB $s$ and user $u$, $FSPL$ is free space path loss in dB, $d_{th}$ is the threshold distance also known as breakpoint distance, $n_1$ and $n_{2}$ is the path loss exponent for distances less than $d_{th}$ and greater than $d_{th}$, respectively.

RSRP of the user is an important performance metric because it gives an estimate of the link strength between the user and the serving gNB. The downlink RSRP $M_u^s$ form the serving gNB $s$ to user $u$ is given by:
\begin{equation}
    M_u^s = P_t^s G_u G_u^s \delta_u^s \Tilde{PL}_{Dual}^{CI}(d_u^s)
\label{eq:RSRP}
\end{equation}
where $P_t^s$ is the transmit power of serving gNB $s$, $G_u$ is the receiver antenna gain of user equipment, $G_u^s$ is the transmitter antenna gain of the serving gNB $s$ towards user $u$, $\delta_u^s$ is the shadowing observed from gNB $s$ at the location of user $u$, $\Tilde{PL}_{Dual}^{CI}(d_u^s)$ is the linear dual slope path loss model derived from eq. \eqref{eq:PL}. $\delta_u^s$ is the shadowing modeled as a gaussian random variable over space.

Since HO inherently impacts cell edge users more than the users in the cell center, it is rational to impart more importance to cell edge users. This motivates us to select edge user RSRP as the first KPI for optimization. The mean RSRP $M$ of the cell edge users in the network can be described as:
\begin{equation}
    M =  \frac{\sum\limits_{\forall i\in \mathbb U} M_s^i}{|\mathbb U|}
\label{eq:RSRP_network}
\end{equation}
where $\mathbb U$ is a set of 25\%-tile RSRP users in the network. The 25\%-tile RSRP provides preference to users with a high chance of HO, i.e., at the cell edge.

Another relevant KPI which is directly affected by HO parameters is the HOSR. The poor setting of HO parameters can lead to low HOSR, which can become a key bottleneck for especially for ultra-reliable low-latency communication (URLLC) use cases. Regardless of the HO type, be it inter- or intra-frequency, HOSR $H$ can be described as:   
\begin{equation}
    H = \frac{HOS}{HOS+HOF} \times 100\%
\label{eq:HOSR}
\end{equation}
where $HOS$ and $HOF$ are the numbers of successful and failed HO, respectively.

Finally, HO parameter setting, inter-frequency HO in particular, can impact the load distribution among different frequency bands. Thus, we incorporate load balancing among different frequency bands as another KPI for efficient resource utilization in the cellular network. Furthermore, balanced data traffic not only provides fairness among users operating at different frequency bands but also minimizes  interference. The load $L_f$ at a frequency band $f$, defined as the average PRB utilization per gNB, can be expressed as:
\begin{equation}
    L_f =  \frac{\sum\limits_{\forall i\in \mathbb B_f} \frac{N^a_i}{N_i}}{|\mathbb B_f|}
\label{eq:load_freq}
\end{equation}
where $N^a_i$ is the number of allocated PRBs at gNB $i$, $N_i$ is the number of all PRBs at gNB $i$ and $B_f$ is a set containing all the gNB of frequency band $f$. The goal of load balancing is to keep a similar PRB utilization across all the frequency bands. We formulate the load balancing among different frequency bands as:
\begin{equation}
    L =  \left(\prod\limits_{\forall i\in \mathbb F}1-L_i\right)^{\frac {1}{|\mathbb F|}} \times100\%
\label{eq:load_band}
\end{equation}
where $\mathbb F$ is a set containing all the frequency bands in the network while we refer $L$ as the load factor for this paper.

\newcounter{storeeqcounter}
\newcounter{tempeqcounter}

\addtocounter{equation}{1}%
\setcounter{storeeqcounter}%
{\value{equation}}%


\begin{figure*}[!t]
\normalsize
\setcounter{tempeqcounter}{\value{equation}} 
\begin{IEEEeqnarray}{rCl}
\setcounter{equation}{\value{storeeqcounter}} 
\begin{array}{clcl}
\displaystyle \max_{A3_{TTT}, A3_{off}, A5_{TTT}, A5_{th1}, A5_{th2}} &  \alpha \Tilde{M} + \beta  \Tilde{H}+ (1-\alpha-\beta) \Tilde{L}; \\
\textrm{subject to} & T1_{min} \leq  A5_{th1} \leq T1_{max}  \\
& T2_{min} \leq  A5_{th2} \leq T2_{max}  \\
& O_{min} \leq  A3_{off} \leq O_{max}  \\
& A5_{TTT}, A3_{TTT}  \in T  \\
& \alpha + \beta \leq 1\\
& L_i \leq L_{th}^i \; \; \; \; \; \; \forall i \in \mathbb F
\end{array}
\label{eq:optimization}
\end{IEEEeqnarray}
\setcounter{equation}{\value{tempeqcounter}} 
\hrulefill
\vspace{-0.2in}
\end{figure*}

We then formulate a multi-objective optimization problem to jointly maximize $M$, $H$, and $L$ employing event A5 related COPs such as $A5_{th1}$, $A5_{th2}$, and $A5_{TTT}$  as well as event A3 related COPs namely $A3_{off}$ and $A3_{TTT}$. The optimization function is given in \eqref{eq:optimization}. $\Tilde{M}$, $\Tilde{H}$ and $\Tilde{L}$ are the normalized values of $M$, $H$ and $L$, respectively. The normalization ensures that the respective weight defines the importance of each KPI by removing the bias towards KPIs with larger values. The operator-defined weight for $M$, $H$, and $L$ are expressed by $\alpha$, $\beta$, and $(1-\alpha-\beta)$, respectively that can be used to adjust their relative importance. To achieve optimal results, the objective function is bounded by several constraints. The first four constraints in \eqref{eq:optimization} limit the values of the optimization variables i.e., COPs in the 3GPP defined ranges. $T1$, $T2$, $O$ are the ranges of $A5_{th1}$, $A5_{th2}$ and $A3_{off}$, respectively, with the subscript showing the minimum and maximum values and $T$ is a set containing all the values of $A5_{TTT}$ and $A3_{TTT}$. Meanwhile, the fifth constraint states that the sum of the three weights should be equal to one, and finally, the last constraint ensures that for each frequency band $i$, the load $L_i$ is less than the operator defined load threshold $L_{th}^i$.


\begin{table}[t]
\centering
\caption{Description of simulation parameters}
\label{table:simParameters}
\begin{tabular}{|l|l|}
\hline
\textbf{Parameter Description}                       & \textbf{Value}                 \\ \hline
Simulation area             & 4km$^2$                     \\ \hline
Number of 1.7GHz macro cells & 6    \\ \hline
Number of 2.1GHz macro cells & 6    \\ \hline
Number of 3.5GHz small cells                      & 12                    \\ \hline
Macro and small cell height & 30m and 20m \\\hline
Macro and cell transmit power     & 35dBm and 20dBm   \\ \hline
Total bandwidth for 1.7, 2.1 and 3.5 GHz       & 10, 15 and 20 MHz                \\ \hline
Total PRBs for 1.7, 2.1 and 3.5 GHz      & 52, 78, 106                \\ \hline
Pathloss exponent 1 and 2 \cite{sun2015path} & 2.9 and 3.9  \\ \hline
Shadowing standard deviation \cite{sun2015path}    &   6.9   \\ \hline
Active user density $\lambda_u$                      & 15 per km$^2$                    \\ \hline
Speed vector $V$                      & [3, 60, 120, 240] km/h                    \\ \hline
Transmission time interval (TTI)   & 1 ms                    \\ \hline
\end{tabular}
\vspace{-0.1in}
\end{table}


\subsection{Simulation Setup and Data Generation}
\label{sec:simSetup}
The data collection from a live network is a tedious task and even impractical in most cases. The reason is the reluctance of most operators to test a wide range of COP combinations due to possible network impairment in a live network in addition to the privacy and business protection concerns. For this reason, data gathered from live networks tend to be sparse and under represented. To address the challenges of sparsity and non-representative data, we leverage a state-of-the-art 3GPP simulator named SyntheticNet \cite{9084113}. To ensure data authenticity, SyntheticNET is calibrated against real network measurements.  



Leveraging SyntheticNet, we deploy a three-tier heterogeneous network in an area of size 2km$\times$2km. The first two layers are composed of tri-sector macro cells operating at 1.7GHz and 2.1GHz frequency bands. On contrary, the third layer consists of omni-directional small cells with 3.5GHz operating frequency. The initial user deployment follows a uniform random distribution with an active user density of $\lambda_u$. The user mobility is modeled as random way point while the speed of each user $v_u$ is chosen from a set $V$ and $v_u$ remains constant. Each entry of $V$ is equally probable. Table \ref{table:simParameters} summarizes the simulation parameters.


The ranges of event A5 and event A3 parameters used to generate the data are shown in Table \ref{table:COPs}. We use a wide range of $A5_{th1}$, $A5_{th2}$, and $A3_{off}$ to cover the impact of hysteresis and make the optimization more robust. To limit the search space size, we use a step size of 2dB for $A5_{th1}$, $A5_{th2}$ and $A3_{off}$. 
A prerequisite step for inter-frequency HO known as measurement gap is implemented using event A2 (i.e., serving RSRP becomes lower than a threshold) with TTT, threshold, and hysteresis set to 64ms, -90dBm, and 1dB, respectively.

\begin{table}[t]
\centering
\caption{Description of COPs to generate the KPIs}
\label{table:COPs}
\begin{tabular}{|>{\centering\arraybackslash} m{2.5cm} | >{\centering\arraybackslash} m{5cm}|}
\hline
\textbf{COPs}                       & \textbf{Values}                 \\ \hline
$A5_{TTT}$      & [64, 128, 192, 256, 384, 512, 640] ms                     \\ \hline
$A5_{th1}$   & [-96 to -116] dBm \\ \hline
$A5_{th2}$    & [-96 to -116] dBm \\ \hline
$A3_{TTT}$    & [64, 128, 192, 256, 384, 512, 640] ms \\ \hline
$A3_{off}$    & [0 to 10] dB \\ \hline
\end{tabular}
\vspace{-0.1in}
\end{table}

\section{Impact of Handover Parameters on KPIs}
\label{sec:impact}
This section presents the impact of varying event A3 and event A5 parameters on the behavior of the selected KPIs. 
The insights from this analysis set the motivation to develop a holistic and concurrent optimization of A3 and A5 parameters. Similarly, the analysis provides valuable insights to network operators for effective tuning of the HO parameters. 

\begin{figure}[t]
\centerline{\fbox{\includegraphics[width=0.48\textwidth]{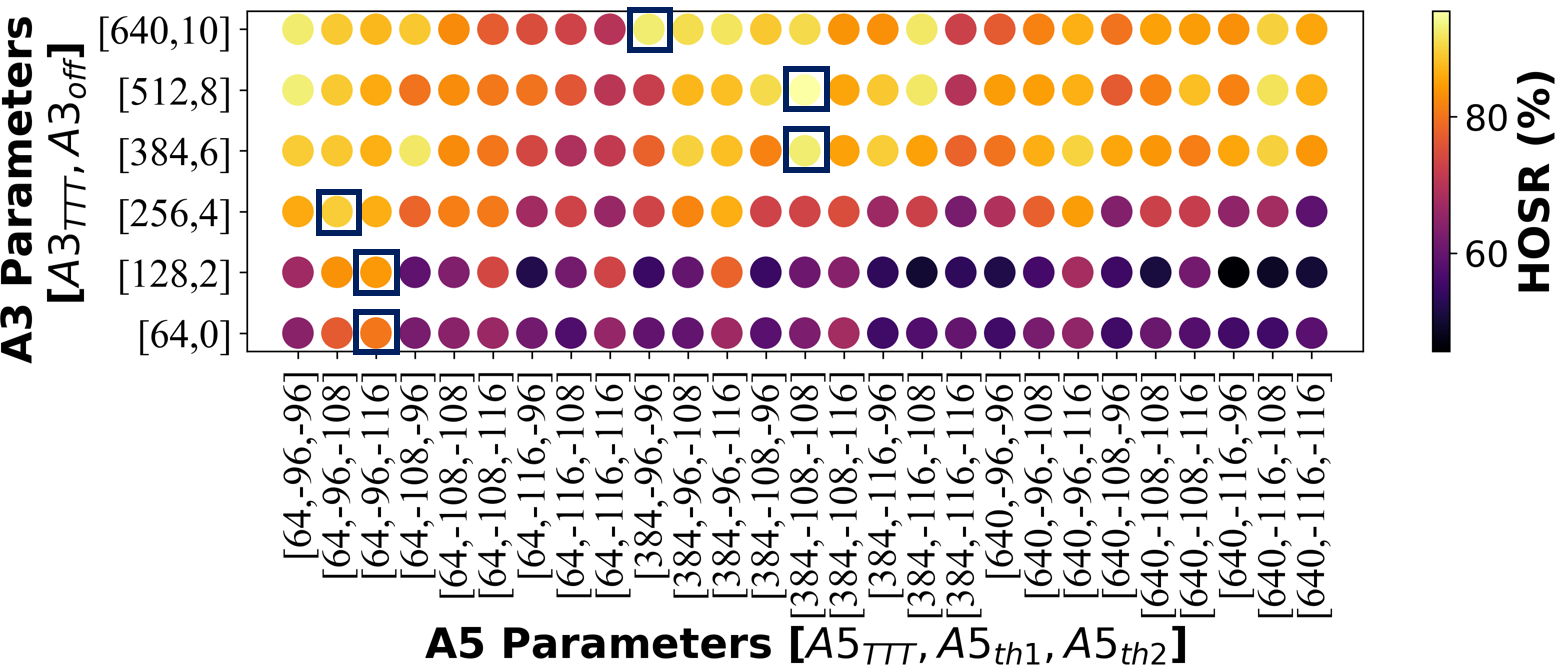}}}
\caption{Impact of A3 and A5 parameters on HOSR. The blue markers highlight the optimal A5 parameters for given A3 parameters.}
\label{fig:HOSR}
\end{figure}

Fig. \ref{fig:HOSR} shows the behavior of HOSR with varying event A3 and event A5 parameters. Analysing event A3 and A5 independently, the first notable observation comes from the apparent increasing trend of HOSR as the values of $A3_{TTT}$ and $A3_{offset}$ become larger regardless of the event A5 parameter setting. A larger value of A3 parameters inherently limits the number of intra-frequency HOs which in turn results to lower chances of HO failures. On the other hand, the trend with varying A5 parameters is not as prominent.    

\begin{figure}[t]
\centerline{\fbox{\includegraphics[width=0.48\textwidth]{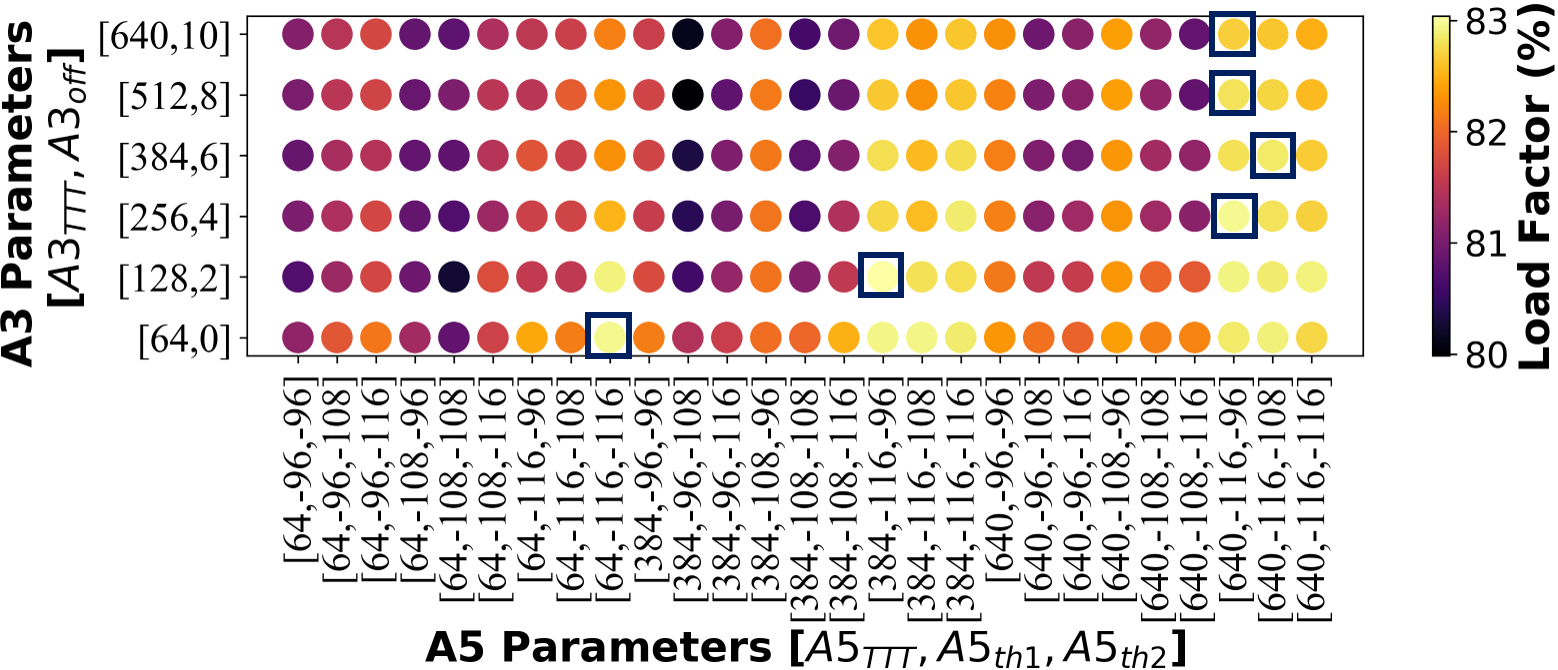}}}
\caption{Impact of A3 and A5 parameters on load factor. The blue markers highlight the optimal A5 parameters for given A3 parameters.}
\label{fig:load_factor}
\end{figure}

Concurrent analysis of event A3 and event A5 reveals that smaller values of A3 parameters combined with larger values of $A5_{th1}$ and smaller values of $A5_{TTT}$ and $A5_{th2}$ provide the optimal HOSR. However, the optimal point shifts to middle values of $A5_{TTT}$, $A5_{th1}$ and $A5_{th2}$ with further increments in the values of A3 parameters. This shift in values of the optimal A5 parameters with variations in A3 parameters signifies an inter-connection between A3 and A5 parameters and optimizing them separately will result in sub-optimal HOSR. This interdependence is further manifested by analysing the impact of varying A3 and A5 parameters on load factor as shown in Fig. \ref{fig:load_factor}. Results reveal that, similar to HOSR, the optimal A5 parameters shift with varying the values of A3 parameters. The results confirm and solidify the need for simultaneous optimization of A3 and A5 parameters.  

Finally, Fig. \ref{fig:HOSR} and Fig. \ref{fig:load_factor} reveal the conflict in optimizing HOSR and load factor. This disparity is highlighted by the difference in the set of A3 and A5 parameters where HOSR and load factor are maximum. Additionally, the results expose the trade-off in optimizing these KPIs. Hence, to balance the trade-off, a multi-objective joint optimization of these KPIs is necessary as proposed in this paper. 




\section{Data-driven Models for Mining COP-KPI Relationship}
\label{sec:ML}
\begin{figure}[t]
\centerline{\fbox{\includegraphics[width=0.48\textwidth]{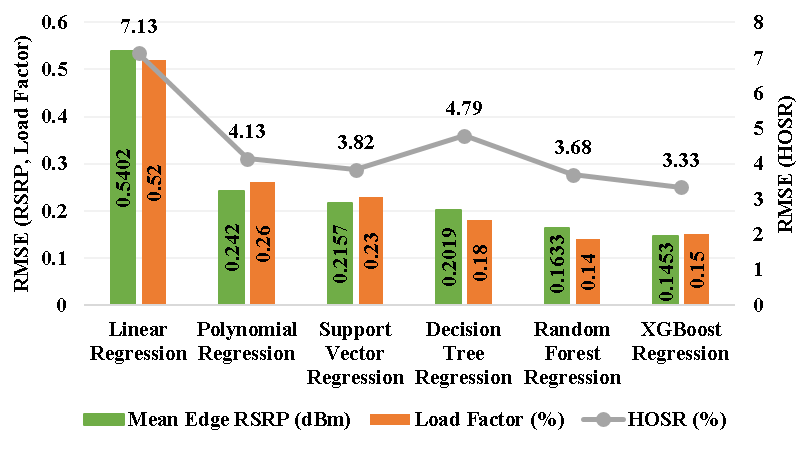}}}
\caption{Comparison of machine learning algorithms in predicting Edge RSRP, Load Factor and HOSR}
\label{fig:ML}
\end{figure}

In this section, we present the performance evaluation of several data-driven models used in mining the relationship between the HO-related COPs and KPIs. Specifically, we leverage six machine learning regression algorithms including linear, polynomial, support vector, decision tree, random forest, and XGBoost. For each KPI, we measure and evaluate the performance of the models in terms of root mean square error (RMSE) using an 80\%-20\% train-test split. Fig. \ref{fig:ML} shows the RMSE of the six algorithms for mean edge user RSRP, load factor, and HOSR. Results reveal that linear regression performs the worst indicating a complex COP-KPI relationship. Furthermore, tree based algorithms generally perform better with decision tree, random forest, and XGBoost outperforming other regression algorithms. XGBoost has the best RMSE of 0.1453dBm and 3.33\% for mean edge RSRP and HOSR, respectively. However, the random forest has the best RMSE of 0.14\% for load factor with XGBoost as a close runner-up.



\begin{figure}[t]
\centerline{\fbox{\includegraphics[width=0.48\textwidth]{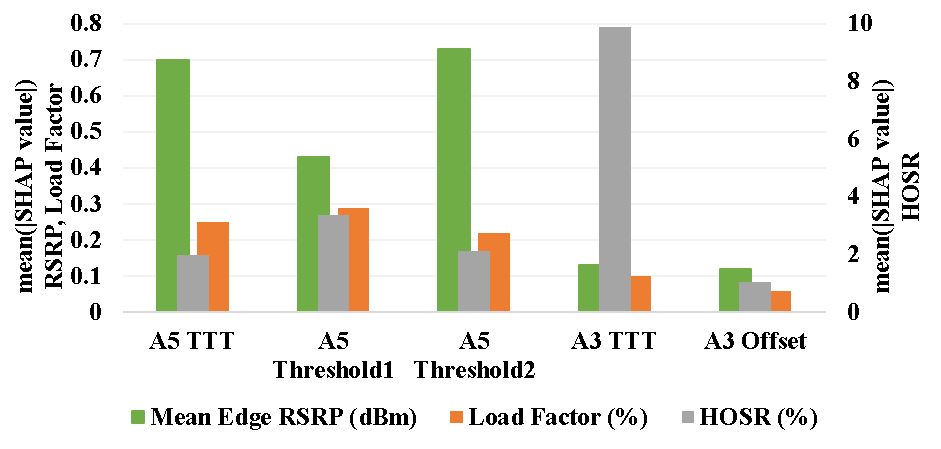}}}
\caption{SHAP sensitivity analysis}
\label{fig:final_SHAP}
\end{figure}

We utilize the SHAP sensitivity analysis \cite{lundberg2017unified} in a bid to get insights into the machine learning models. Fig. \ref{fig:final_SHAP} shows the average impact of all the five COPs on mean edge user RSRP, load factor, and HOSR. Results show that $A5_{TTT}$ and $A5_{th2}$ have the highest impact on mean edge RSRP while $A3_{TTT}$ and $A3_{off}$ have minimal impact. On the other hand, $A5_{TTT}$ and $A5_{th1}$ influence load factor the most followed by $A5_{th2}$. Furthermore, the load factor is least impacted by $A3_{TTT}$ and $A3_{off}$. In contrast to the edge user RSRP and load factor, $A3_{TTT}$ has the highest mean SHAP values for HOSR while $A3_{off}$ has the least impact. This SHAP analysis reveals that the relative impact of each COP on the three KPIs is different. For instance, $A3_{TTT}$ has a very high impact on HOSR compared to mean edge RSRP and load factor. Hence, $A3_{TTT}$ can be tuned for optimizing HOSR without significant degradation in other KPIs. These SHAP-based valuable insights can be leveraged for multi-objective parameter tuning by network operators and can be leveraged to avoid existing SON conflicts. 

\begin{figure}[t]
\centerline{\fbox{\includegraphics[width=0.48\textwidth,height=2.1in]{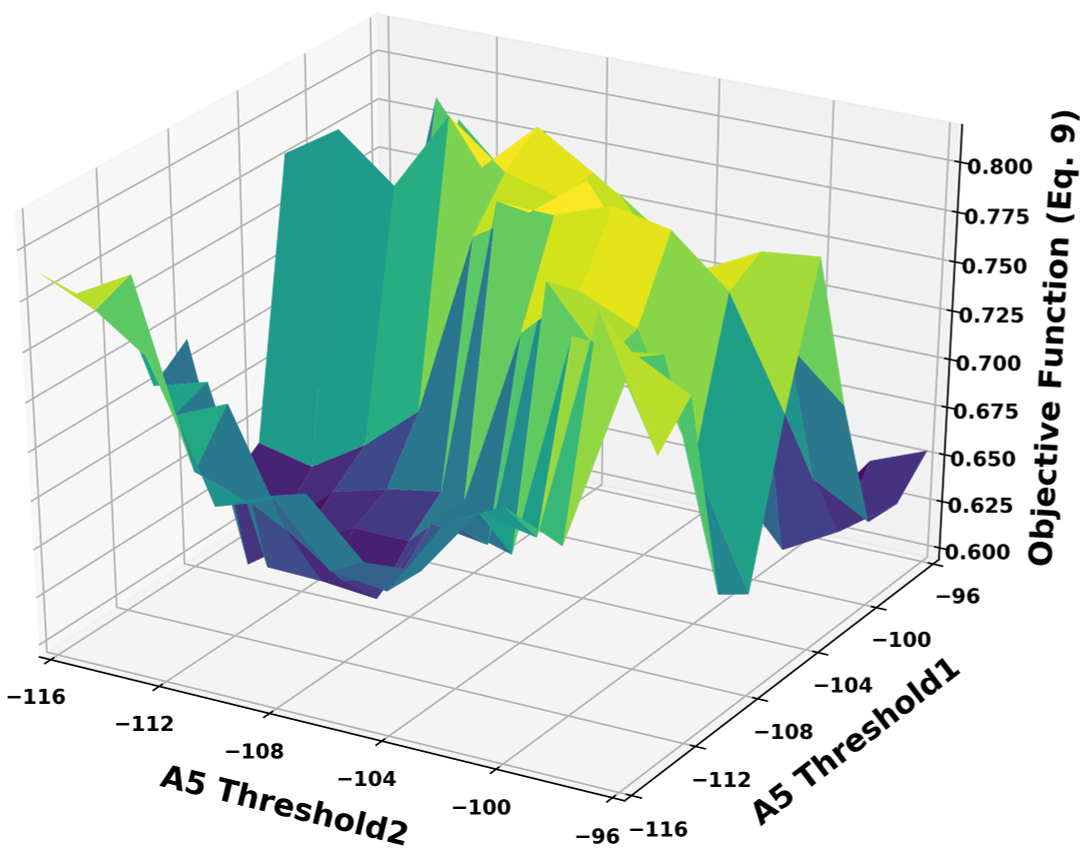}}}
\caption{Objective function defined in eq. \eqref{eq:optimization} with $A5_{TTT}=64$, $A3_{TTT}=64$, $A3_{off}=0$, $\alpha=0.33$ and $\beta=0.33$}
\label{fig:obj_func}
\end{figure}

\section{Objective Function Optimization}
\label{sec:optimization}

The objective function defined in eq. \eqref{eq:optimization} with $A5_{TTT}=64$, $A3_{TTT}=64$, $A3_{off}=0$, $\alpha=0.33$ and $\beta=0.33$ is shown in Fig. \ref{fig:obj_func}. The presence of multiple maxima in the plot reveals the non-convex nature of the optimization problem. To solve this types of optimization problems, usual approaches include brute force or heuristic optimization solutions. In this paper, we compare the performance of the brute force and SA in solving eq. \ref{eq:optimization}. Although SA does not always converge to the best solution, its convergence time is significantly lesser than brute force. Table \ref{table:SA} shows that simulated annealing converges more than 14 times faster than brute force. The quick convergence of simulated annealing makes the solution agile, which is particularly important for rapidly changing network conditions.

A comparison of the best objective function obtained by SA+XGBoost and brute force is shown in Table \ref{table:SA}. We analyse the resulting objective function for different values of $\alpha$ and $\beta$ to test the robustness of the proposed solution with varying importance of KPIs in the objective function. We implement a monte-carlo simulation for 1000 iterations for SA. Results show a small difference in the objective function obtained using SA+XGBoost and brute force for different KPI importance. This small difference in the objective function is due to the prediction error in the ML algorithms and slight non-optimal convergence of SA. Moreover, these results indicate that the proposed ML-aided SA solution can enable agile self optimization of mobility parameters.


\begin{table} [t]
\centering
\caption{{Comparison between Simulated Annealing and Brute Force Approach}}
\label{table:SA}
\begin{tabular}{|>{\centering\arraybackslash} m{1.3cm} |>{\centering\arraybackslash} m{0.4cm} | >{\centering\arraybackslash} m{0.4cm} | >{\centering\arraybackslash} m{1.9cm} |  >{\centering\arraybackslash} m{0.9cm}|}
\hline
& $\bm{\alpha}$ & $\bm{\beta}$ & \textbf{SA+XGBoost} &  \textbf{Brute Force} \\ \hline
\multirow{4}{*}{\shortstack{\textbf{Objective}\\ \textbf{Function}}} & 0.33 & 0.33 & 0.8684 & 0.8878 \\ \cline{2-5}
& 0.8 & 0.1 & 0.9380 & 0.9610 \\ \cline{2-5}
& 0.1 & 0.8 & 0.9096 & 0.9375  \\ \cline{2-5}
& 0.1 & 0.1 & 0.9011 & 0.9414  \\ \hline
\multicolumn{3}{|c|}{\textbf{Number of Iterations}} & 2500 & 35574  \\ \hline 
\end{tabular}
\vspace{-0.1in}
\end{table}

\section{Conclusion}
\label{sec:conc}
Network densification and multi-band network deployment pose new challenges to HO management. In this paper, we present a solution that concurrently optimizes inter-frequency and intra-frequency HO parameters to jointly maximize mean edge user RSRP, load factor, and HOSR. We leverage machine learning to quantify the COP-KPI relationship as tractable analytical solutions are not possible due to the system level complexity. The evaluation shows that XGBoost performs the best in capturing the behavior of HOSR and mean edge user RSRP with varying HO parameters while random forest has the best performance for load factor. SHAP sensitivity analysis reveals that event A5-related parameters are the most important COPs for mean edge user RSRP and load factor. On the other hand, the event A3-related parameter specifically TTT has the highest importance for HOSR. After showing the non-convex behavior of the objective function, we evaluate the performance of the brute force and simulated annealing using different KPI priority weights. Results show that the ML-based SA aided solution is more than 14 times faster compared to brute force at the cost of slight sub-optimal objective function.


\section*{Acknowledgment}
This work is supported by the National Science Foundation under Grant Numbers 1718956 and 1730650, and Qatar National Research Fund (QNRF) under Grant No. NPRP12-S 0311-190302. The statements made herein are solely the responsibility of the authors. For more details about these projects please visit: http://www.ai4networks.com

\ifCLASSOPTIONcaptionsoff
  \newpage
\fi

\bibliographystyle{ieeetr}
\bibliography{IoT}

\end{document}